\documentclass[aps,prl,twocolumn,groupedaddress]{revtex4}
\usepackage{graphicx}

\begin{document}


\title{Luttinger liquid of polarons in one-dimensional
boson-fermion mixtures}


\author{L. Mathey, D.-W. Wang, W. Hofstetter, M. D. Lukin,
and Eugene Demler}
\affiliation{Physics Department, Harvard University, Cambridge, MA 02138}


\date{\today}

\begin{abstract}
We use bosonization approach to investigate 
quantum phases in mixtures of bosonic and fermionic atoms confined in one
dimensional optical lattices. The phase diagrams can be  well understood in terms of polarons, which correspond to  atoms that are "dressed"  by screening clouds of the other atom species.   For a mixture of single species of 
fermionic and bosonic atoms we find a charge density
wave phase, a phase with fermion pairing, and a regime of phase
separation. For a mixture of two species of fermionic atoms and one
species of bosonic atoms we obtain spin and charge density wave
phases, a Wigner crystal phase, singlet and triplet paired states of
fermions, and a phase separation regime.   Equivalence
between the Luttinger liquid description of polarons and the canonical polaron transformation
is established and the techniques to detect the resulting quantum 
phases are discussed. 
\end{abstract}


\maketitle
Mixtures of ultra-cold bosonic and fermionic atoms, that have recently become accessible experimentally, represent  
a promising new system for studying strongly correlated many-body
physics \cite{bfm}. Bosonic atoms mediate
interactions between fermions and allow efficient cooling of the
system \cite{sympathetic_cooling}. Several novel phenomena have
been predicted theoretically for Boson-Fermion mixtures (BFM)
including pairing of fermions \cite{pairing}, formation of composite
particles \cite{composite}, spontaneous breaking of translational
symmetry in optical lattices \cite{burnett} and 
appearance of charge density wave
(CDW) \cite{xCDW}. Most of these theoretical studies  relied on
integrating out bosonic degrees of freedom to obtain an effective interaction between
fermions, and then using a mean-field approach to investigate many-body
states \cite{pairing}. 
This approach, however, becomes unreliable in the regime of 
strong interactions. In particular, it  fails in low-dimensional systems due to enhanced fluctuations and non-perturbative effects of interactions.

\begin{figure}
\includegraphics[width=3.5cm]{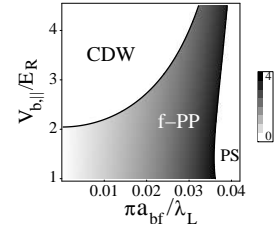}
\caption{Phase diagram for a mixture of bosonic and spinless
fermionic atoms in a 1D optical lattice. 
Shading in the $f$-PP phase describes the strength
of the bosonic screening cloud ($2 \lambda$, see Eq.(\ref{polaron})) around a pair of fermions. $\lambda_L$ and $E_R$ are respectively the lattice period
and recoil energy.
Other parameters used for this figure are (see text for notations): 
$\nu_b=4$, $\nu_f=0.5$,
$V_{b,\perp}=V_{f,\perp}=20 E_R$, $V_{f,\|}=2 E_R$,  boson-boson 
scattering length $a_{bb}=0.01 \lambda_L$.
}
\label{phase_diagRP}
\end{figure}

\begin{figure}
\includegraphics[width=8cm]{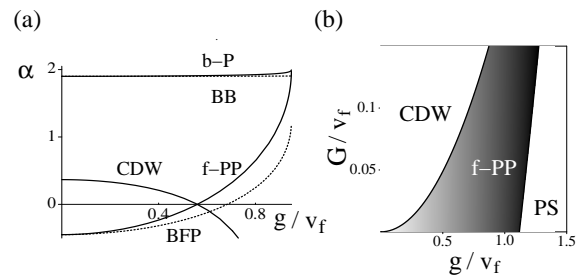}
\caption{Ground state of a BFM with spinless fermions.
$g$ is the longitudinal FP coupling and $G$ is the effective
fermion-fermion interaction after integrating out $2k_f$ phonons.
(a) Scaling exponents for $v_b/v_f = 3$, $K_b = 5$ and $G/v_f = 0.1$.
Different curves correspond to the $2k_f$ CDW order
parameter, $f$-polaron pairing field ($f$-PP), bare fermion pairing
field (BFP), $b$-polaron operator ($b$-P), and bare boson operator
(BB). Note that operators
constructed with polarons (bosons and Cooper pairs of
fermions) always have larger exponents than their counterparts
constructed with bare atoms. (b) Global phase diagram for $v_b/v_f=5$
and $K_b=10$.}
\label{phase_diag_alpha}
\end{figure}

\begin{figure}
\includegraphics[width=8cm]{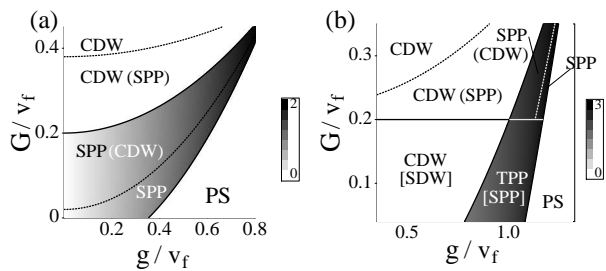}
\caption{
Phase diagrams for a mixture of bosonic and $S=1/2$ fermionic 
atoms. Shading in the paired regions describes the strength
of bosonic screening clouds around fermionic
polarons. In both figures $v_b/v_f=5$ and $K_b=10$. 
In (a) $U_{\uparrow\downarrow}/v_f=-0.8\pi$
and in (b) $U_{\uparrow\downarrow}/v_f=0.8\pi$.
At the top of (b) and everywhere in (a) we have
$g_{1\perp}<0$ ($K_{\sigma}<1$), and $g_{1\perp}<0$ ($K_{\sigma}>1$) at the bottom of (b). 
For larger $G$ and $U_{\uparrow\downarrow}$ one can also have a Wigner crystal phase (not shown here). Note that 
parentheses ($\dots$) indicate subdominant phases, while square 
brackets [$\dots$] indicate degenerate phases. 
}
\label{phase_diag_spin}
\end{figure}

In this paper we use bosonization method 
\cite{haldane_bf,cazalilla} to investigate one dimensional (1D) BFM. 
The resulting quantum phases can be understood by introducing 
polarons,
 i.e. atoms of one species
surrounded by screening clouds of the other species. Such dressed quasi-particles
exhibit effective interactions and modified effective masses. In our 
analysis the polarons  emerge as quasi-particles with the slowest decaying correlation functions while quantum phases of the system arise from a competition of various ordering instabilities of such polarons. 
The phase diagrams we obtain (Figs. \ref{phase_diagRP}-\ref{phase_diag_spin}) show a
remarkable similarity to 
the Luttinger liquid phase diagrams of 1D {\it interacting} electron systems
\cite{solyom}, suggesting that 1D BFM may be understood as Luttinger liquids
of polarons.

In Fig. \ref{phase_diagRP} we show a
phase diagram  for a mixture of bosons and spinless fermions
as a function of experimentally controlled parameters: the scattering
length between bosons and fermions ($a_{bf}$) and the strength
of the longitudinal optical lattice for bosonic atoms ($V_{b,\|}$)
\cite{endnote1}.
For relatively weak boson-fermion interactions and slow bosons (i.e. 
strong optical lattice for bosonic atoms) the system is in the CDW phase, in which the densities of fermions and bosons have a $2 k_f$-modulation.
In the case of very strong boson-fermion interactions the system is unstable to phase
separation (PS) \cite{ho,instability}. 
The two regimes are separated by a p-wave pairing phase of fermionic polarons  ($f$-PP).
While we carry out the detailed analysis for atoms in optical lattices, 
qualitatively it also applies to the continuous
case  \cite{bfm,sympathetic_cooling,general_1D}.

Before proceeding we note that bosonization  approach has been applied to BFM in Ref.[\onlinecite{ho}].  However this work did not consider correlations of polaronic degrees of freedom, and as a result 
did not predict most of the quantum phases. 
The present system also  resembles  1D electron-phonon systems discussed previously in
Ref.[\onlinecite{voit_phonon}]. The key difference between the two is that the sound velocity in a solid state is typically smaller than the Fermi velocity, whereas for the dilute fermionic gas coupled to superfluid
bosonic ensemble the opposite is true. In particular, this rules out an adiabatic elimination of the bosonic field. 
In contrast, we demonstrate in the present system polaronic
quasi-particles dominate the low-temperature behaviour of the system. 
We also note that the 1-D $p$-wave superfluid we obtain here 
may be of relevance to recent studies in quantum information in Ref.[\onlinecite{kitaev}].

We first consider a mixture of spinless fermionic
($f$) and bosonic ($b$) atoms. For sufficiently strong optical potential the
microscopic Hamiltonian is given by a single band Hubbard model
\begin{eqnarray}
H &=& - \sum_{\langle ij\rangle} \left( t_b b_i^{\dagger} b_j 
+ t_f f_i^{\dagger} f_j \right)
-\sum_i\left(\mu_f n_{f,i}+\mu_b n_{b,i}\right)
\nonumber\\
&&+ \frac{U_{b}}{2} \sum_i  n_{b, i} (n_{b,i} - 1)
+U_{bf} \sum_i n_{b, i} n_{f, i},
\label{H_tot}
\end{eqnarray}
Here $n_{b,i}\equiv b^\dagger_i b^{}_i$ and $n_{f,i}
\equiv f^\dagger_i f^{}_i$ are respectively
the boson and fermion density operators, and $\mu_{b/f}$
are their chemical potentials.  The tunneling amplitudes $t_{f/b}$,
and the particle interactions $U_{b}$ and 
$U_{bf}$ depend on the laser beam intensities and the $s$-wave scattering
lengths, and can be calculated explicitly (see e.g. 
Ref.[\onlinecite{Jaksch98}]). 
In this paper we assume that the filling
fraction of fermions  $\nu_{f}\equiv\langle
n_{f,i}\rangle$ is not commensurate with the lattice
or with the filling fraction of bosons $\nu_{b}$. 
The  Fermi momentum and velocity are given by $k_f=\pi\nu_f$ 
and $v_f=2t_f\sin(k_f)$,respectively.

The Haldane's bosonization representations
for fermion and boson operators are constructed as \cite{haldane_bf,ho}
$f(x)=\left[\nu_f+\Pi_f\right]^{1/2}\sum_{m=-\infty}^{\infty}
e^{(2m+1)i\Theta_f}e^{i\Phi_f}$ and
$b(x)=\left[\nu_b+\Pi_b\right]^{1/2}\sum_{m=-\infty}^{\infty}
e^{2mi\Theta_b}e^{i\Phi_b}$,
where $x$ is a continuous coordinate that replaces the site index $i$
in Eq. (\ref{H_tot}). The operators
$\Pi_{f/b}(x)$ and $\Phi_{f/b}(x)$ describe the bosonized
density and phase operators for each of the atomic species, and
satisfy the commutation relations
$\left[\Pi_{b/f}(x),\Phi_{b/f}(x')\right]=i\delta(x-x')$.
Short-ranged fluctuations are taken into account
via the harmonics of 
$\Theta_{f/b}(x)=\pi\nu_{f/b}x+\pi\int^x dy\,\Pi_{f/b}(y)$.
The low energy effective Hamiltonian can be written to be the sum of
the following terms:
\begin{eqnarray}
H_f^0 &=& \frac{v_f}{2}\int_0^L dx\left[\frac{1}{\pi}
\left(\partial_x\Phi_f(x)\right)^2+\pi\Pi_f(x)^2\right]
\label{H_f^0}
\\
H_b^0 &=& \frac{v_b}{2}\int_0^L dx\left[
\frac{K_b}{\pi}\left(\partial_x\Phi_b(x)\right)^2+
\frac{\pi}{K_b}\Pi_b(x)^2\right]
\label{H_b^0}
\\
H^1_{bf} &=& U_{bf}\int_0^L dx\, \Pi_b(x)\Pi_f(x)
\label{H_bf^1}
\\
H^2_{bf} &=& U_{bf}\int_0^L dx\,\Pi_b(x)\tilde{n}_f(x),
\label{H_bf^2}
\end{eqnarray}
where we neglected the Umklapp scattering
and backward scattering of bosons by assuming that bosons
are deep inside the superfluid regime \cite{halffilling}. When the
boson-boson interaction is weak (i.e. $\gamma\equiv m^\ast_b U_b/\nu_s<
10$), the phonon velocity $v_b$ and the
scaling exponent of bosons $K_b$ in Eq.(\ref{H_b^0}) can be well
approximated by $v_b =
\frac{\nu_s}{m^\ast_b}\sqrt{\gamma}\left(1-\frac{\sqrt{\gamma}}{2\pi}
\right)^{1/2}$ and $K_b =
\frac{\pi}{\sqrt{\gamma}}\left(1-\frac{\sqrt{\gamma}}{2\pi}
\right)^{-1/2}$, where $m^\ast_b$ is the effective boson mass in the
presence of the lattice potential \cite{stringari} and $\nu_s \sim \nu_b$ is the
superfluid fraction. The term $H_{bf}^1$ in Eq. (\ref{H_bf^1}) describes forward
scattering between bosons and fermions, and the term $H_{bf}^2$
corresponds to back scattering of fermions on bosons (We defined
$\tilde{n}_f(x)=\frac{1}{L}\sum_{k\sim 2k_f} \sum_{p\sim k_f}\left[
e^{ik x} f^\dagger_{R,-p+k}f_{L,-p}^{} +e^{-ik
x}f^\dagger_{L,p-k}f_{R,p}^{}\right]$, where $f_{\{ R,L \}
,p}^\dagger$ are the creation operators for the right and left moving
fermions near the Fermi energy).  
%
%
It is useful,
however, to take into account the deviations of the phonon dispersion from
the linear spectrum by using the Bogoliubov
approximation $\omega_{k}=\sqrt{(\varepsilon_{b,k}-\varepsilon_{b,0})
(\varepsilon_{b,k}-\varepsilon_{b,0}+2U_b\nu_b)}$, where
$\varepsilon_{b,k}$ is the band energy of a noninteracting boson. 

We integrate out the $2 k_f$-bosons and obtain an effective fermion-fermion interaction.
Since we are considering a BFM with fast phonon velocity, i. e. $v_b \gg v_f$, we obtain within instantaneous approximation:
%
%
\begin{eqnarray}
H_f^{1}&=&\frac{2G}{2\pi}\int_0^L
dx\left[\pi^2\Pi_f(x)^2-(\partial_x\Phi_f(x))^2\right].
\label{H_f^1}
\end{eqnarray}
Here $G\equiv\frac{g_{2k_f}^2}{\omega_{2k_f}}$ is the induced 
fermion-fermion interaction, and 
$g_k$ is the fermion-phonon (FP) coupling vertex,
$g_k=U_{bf}\sqrt{\nu_b(\varepsilon_{b,k}-\varepsilon_{b,0})/2\pi\omega_k}$.
Note, that for small $k$ we have a conventional
FP coupling $g_k=g |k|^{1/2}$ with $g=U_{bf}\sqrt{K_b}/2\pi$.
Therefore the effective Hamiltonian for a BFM is given by
Eqs. (\ref{H_f^0})-(\ref{H_bf^1}) and
(\ref{H_f^1}) with five parameters: 
$v_f$, $v_b$, $K_b$, $g$ and $G$. 
We can diagonalize this Hamiltonian \cite{Engelsberg}
%
%
%
%
and obtain
\begin{eqnarray}
H&=&\frac{1}{2}\sum_{j=a,A}v_j\int dx\left[\pi\Pi_j(x)^2
+\frac{1}{\pi}\left(\partial_x\Phi_j(x)\right)^2\right]
\label{new_hamiltonian}
\end{eqnarray}
where the eigenmode velocities, $v_A$ and $v_a$, are given by
\begin{eqnarray}
v_{a/A}^2 & = & \frac{1}{2} (v_b^2 + \tilde{v}_f^2) \pm \frac{1}{2} 
\sqrt{(v_b^2 - \tilde{v}_f^2)^2 +  16 \tilde{g}^2  v_b \tilde{v}_f}.
\label{new_velocity}
\end{eqnarray}
Here $\tilde{v}_f\equiv(v_f^2-4G)^{1/2}$ and $\tilde{g}\equiv
g\,e^\theta$ with $e^\theta = 
((v_f - 2 G)/(v_f + 2 G))^{1/4}$.  

When the FP coupling $g$
becomes sufficiently strong, the eigenmode
velocity $v_A$ becomes soft, indicating an 
instability of the system to phase separation or
collapse, depending on the sign of $U_{bf}$ \cite{ho,instability}.
To understand the nature of the many-body state of BFM 
outside of the instability region
we analyze the long distance behavior of the correlation functions.
For the bare bosonic and fermionic particles we find $\langle
b(x)b^\dagger(0)\rangle \sim |x|^{-\frac{1}{2}K_\epsilon^{-1}}$ and
$\langle f(x)f^\dagger(0)\rangle \sim \cos(k_f x)
|x|^{-\frac{1}{2}(K_\beta+K_\gamma^{-1})}$
(see Ref.[\onlinecite{Definitions}]). To describe particles dressed
by the other species
we introduce the composite operators
\begin{eqnarray}
\tilde{f}_\lambda(x) \equiv e^{-i\lambda\Phi_b(x)} f(x),
\ \ \ 
\tilde{b}_\eta(x) \equiv e^{-i\eta\Phi_f(x)} b(x),
\label{polaron}
\end{eqnarray}
with $\lambda$ and $\eta$ being real numbers. The
correlation functions of these operators
are given by $\langle
\tilde{f}(x)\tilde{f}^\dagger(0)\rangle \sim \cos(k_f x)
|x|^{-\frac{1}{2}(K_\beta+\lambda^2 K_\epsilon^{-1} +K_\gamma^{-1}
-2\lambda K_{\gamma\epsilon}^{-1})}$ and $\langle
\tilde{b}(x)\tilde{b}^\dagger(0)\rangle \sim
|x|^{-\frac{1}{2}(K_\epsilon^{-1}+\eta^2 K_\gamma^{-1} -2\eta
K_{\gamma\epsilon}^{-1})}$ (see Ref.[\onlinecite{Definitions}]).  
We observe that the exponents of the
correlation functions are maximized with
$\lambda_c=K_\epsilon/K_{\gamma\epsilon}$ and
$\eta_c=K_\gamma/K_{\gamma\epsilon}$, which we use
to construct polaronic particles according to Eq. [\ref{polaron}]. In the limit of weak
interactions we have $\lambda_c\to U_{bf}/U_b$ and
$\eta_c\to 2 U_{bf}/\pi v_b$. This result can be understood
by a naive density counting argument
that a fermionic polaron ($f$-polaron)
locally suppresses a bosonic cloud by
$\lambda_c$ particles, whereas a bosonic polaron ($b$-polaron)
depletes the fermionic system by $\eta_c$ atoms.

Ground states of one dimensional systems are often characterized by
specifying the order parameters that have the slowest long distance decay
of the correlation functions \cite{solyom}. This is equivalent to
finding the most divergent susceptibility in the low temperature limit
(When the $T=0$ correlation function decays as $\langle O(x) O(0)
\rangle \sim 1/|x|^{2-\alpha}$, the finite $T$ susceptibility diverges
as $\chi(T) \sim 1/T^{\alpha}$). For the $2k_f$ CDW order
parameter, $O_{CDW}=f^\dagger_{L} f_{R}$, we find $\alpha_{CDW}= 2-2K_\beta$,
and for the $f$-polaron pairing field,
$O_{f-PP}=\tilde{f}_{L\lambda}\tilde{f}_{R\lambda}$, we obtain
$\alpha_{f-PP}=2-2\left[\lambda^2K_\epsilon^{-1}
+K_\gamma^{-1}-2\lambda K_{\gamma\epsilon}^{-1}\right]$.
We did not include
polaron dressing in $O_{CDW}$, since this operator has no net
fermionic charge and the exponent
of $O_{CDW}$ does not change if we replace $f$ by $\tilde{f}$.
For the $f$-PP operator, on the other hand, $\alpha_{f-PP}$ is maximized
when $\lambda=\lambda_c$, so from now on $\alpha_{f-PP}$ will always
mean the $f$-PP exponent computed with the optimal $\lambda_c$. Scaling
exponents shown in Fig. \ref{phase_diag_alpha}(a) demonstrate that
divergencies of the CDW and $f$-PP susceptibilities are mutually exclusive
and cover the entire phase diagram. In Fig.
\ref{phase_diag_alpha}(b) we show a global phase diagram of a BFM
considering the FP coupling ($g$) and effective fermion-fermion interaction
($G$) as independent variables. This phase diagram is
similar to what one finds for spinless electrons in Luttinger liquid
theory \cite{solyom}. The phase diagram in terms of experimentally controlled parameters was shown 
in Fig. \ref{phase_diagRP}. We point out that if we were to define the pairing
operator using bare fermions rather than $f$-polarons with optimal
$\lambda_c$, we would have a region in the phase diagram in which
none of the susceptibilities diverges (for the parameters used in
Fig. \ref{phase_diag_alpha}(a) this regime extends between
$0.55<g/v_f<0.68$). 

We next verify that the conventional construction of polaron
operators based on the canonical polaron transformation (CPT)
\cite{mott} can give an equivalent polaronic description as in Eq.[\ref{polaron}]. The  CPT operator is given by
$U_\lambda=e^{-i\lambda\sum_{{k}\neq 0}\left(
F_{k}\beta_{k}\rho_{k}^\dagger+{\rm h.c.}\right)}$, where $\beta_{k}$
is the phonon annihilation operator, $\rho_{k}$ is the fermion density
operator, $F_{k}$ is some function of momentum ${k}$, and $\lambda$
specifies the strength of the phonon dressing. When applied to a
fermion operator, the CPT gives $U^{-1}_\lambda f(x)
U^{}_\lambda=f(x)\exp\left[-i\lambda\sum_{{k}\neq 0}
\left(F_{k}\beta_{k} e^{-i{k}\cdot{x}}+{\rm h.c.}  \right)\right]$
\cite{mott}, which is the same as Eq.(\ref{polaron}), provided
that one takes $F_k=\frac{1}{2}\sqrt{\frac{2\pi}{K_b|k|L}}{\rm
sgn}(k)$ (note that in 1D fermionic systems
density operators correspond to Luttinger bosons).

At finite temperature the correlation functions become $\langle O(x) O(0)
\rangle \sim \exp(-|x|/\xi)/|x|^{2-\alpha}$. The thermal correlation lengths for $O_{CDW}$ and $O_{f-PP}$ are approximately given by $\xi \sim v_f /k_B T$. For a finite system of length $L=N \lambda_L$ ($N$ being the number of lattice sites), the $T=0$-properties of the system are visible for $\xi \sim L$. This corresponds to a temperature regime of $T \sim T_f/N$, $T_f$ being the Fermi temperature.

Before concluding our discussion of BFM with spinless fermions we
discuss an approach for observing the phase transition between
the CDW and the $f$-PP phases in laser stirring experiments \cite{stirring}. Suppose that a laser beam is focused at the center of the cloud, such that it 
creates a weak local potential for fermionic atoms.  In the $f$-PP phase the
stirring potential can be moved through the system with no
dissipation, if its velocity is slower than some critical
value \cite{stirring}. At the $f$-PP/CDW phase
boundary the critical velocity goes to zero, reflecting a transition
to the insulating (CDW) state.  This scenario follows from an RG analysis of a small local potential for fermions: Such a potential is
irrelevant in the $f$-PP phase, but becomes relevant in
the CDW phase, similar to the single impurity
problem of a 1D electron system \cite{kane}. It should be emphasized that only when polarons are used to
construct Cooper pairs, the paired phase coincides with the domain of
irrelevance of a small pinning potential.

We now extend our analysis to BFM with fermions with two internal
hyperfine states, which we assume to be $SU(2)$ symmetric.  Spin
symmetry of the system leads to separation of the bosonized
Hamiltonian into spin and charge sectors.  The charge part of the
Hamiltonian is equivalent to a BFM with spinless atoms and can be diagonalized analogously. The spin part of the
Hamiltonian has a form of the sine-Gordon model, $H_\sigma =
\frac{1}{2}v_\sigma\int dx\left[\pi\Pi_\sigma(x)^2
+\frac{1}{\pi}\left(\partial_x\Phi_\sigma(x)\right)^2\right] +\frac{2
g_{1\perp}}{(2\pi\alpha)^2} \int
dx\cos\left[\sqrt{8K_\sigma}\Theta_\sigma(x)\right],$ where
$v_\sigma$ is the spin velocity, $K_\sigma$ is the spin Luttinger
exponent, and $g_{1\perp}=U_{\uparrow\downarrow} -4\pi G$ is the
effective backward scattering amplitude for fermions, that has
contributions from the bare fermion-fermion interaction and from integrating
out $2k_f$ phonons.  The nature of spin excitations in
the ground state follows from the well known properties of the
sine-Gordon model: For $K_{\sigma} > 1$ the system has
gapless spin excitations ($g_{1\perp}$ is irrelevant), for $K_{\sigma} < 1$ the system has a spin gap
($g_{1\perp}$ becomes relevant).  In
order to describe possible ground states of the system we calculated
the low temperature behavior of susceptibilities for the following
order parameters: $2k_f$ spin density wave ($\chi_{SDW}$), $2k_f$
charge density wave ($\chi_{CDW}$), $4k_f$ (Wigner crystal) charge
density wave ($\chi_{WC}$), singlet polaron pairing ($\chi_{SPP}$), and
triplet polaron pairing ($\chi_{TPP}$). Depending on the parameters we
find the following regimes: 1) Equally divergent $\chi_{CDW}$ and
$\chi_{SDW}$.  Degenerate CDW and SDW phases; 2) Equally divergent
$\chi_{SPP}$ and $\chi_{TPP}$.  Degenerate SPP and TPP phases; 3)
Divergent $\chi_{SPP}$. State with singlet pairing of polarons; 4)
Both $\chi_{CDW}$ and $\chi_{SPP}$ diverge, but unequally. This probably
corresponds to a phase that has both a CDW order and singlet polaron
pairing; 5) Divergent $\chi_{CDW}$. CDW phase; 6) Divergent
$\chi_{WC}$. Wigner crystal phase. The phase diagram 
in Fig. \ref{phase_diag_spin} again 
shows remarkable similarity to a phase diagram for interacting
electrons \cite{solyom}.

In summary, we used the bosonization method to
investigate 1D mixtures of bosonic and fermionic atoms involving spinless
and $S=1/2$ fermions. Interactions between atoms can lead to such
interesting phenomena as spin and charge density waves, singlet and
triplet pairing of atoms.  This phase diagram can be understood 
in terms of polarons, in that it corresponds to a  Luttinger liquid of polarons.
We also discussed laser stirring experiments as a technique for
probing quantum phase transitions between paired and insulating phases
of polarons, and considered the finite temperature and size of a real system.

We thank B.I. Halperin for useful discussions.
This work was supported by the NSF (grants DMR-01328074, PHY-0134776),
the Sloan and the Packard Foundations, and by Harvard-MIT CUA. W.H. was supported by the German Science Foundation (DFG) and by the NSF (grant DMR-02-33773). 


\begin{thebibliography}{99}
\bibitem{bfm}
G. Modugno, \emph{et al.}, Science {\bf 297}, 2240 (2002);
F. Ferlaino \emph{et al.}, cond-mat/0312255; 
A. Simoni \emph{et al.}, cond-mat/0301159;
B. Laburthe \emph{et al.}, cond-mat/0312003;
T. St\"{o}ferle \emph{et al.}, cond-mat/0312440.

\bibitem{sympathetic_cooling}
Z. Hadzibabic \emph{et al.}, Phys. Rev. Lett. {\bf 88}, 160401 (2002);
G. Roati, \emph{et al.}, Phys. Rev. Lett. {\bf 89}, 150403 (2002);
F. Schreck, {\it et al.}, Phys. Rev. Lett. {\bf 87}, 080403 (2001);
A.G. Truscott, Science {\bf 291}, 2570 (2001).

\bibitem{pairing}
F. Matera, Phys. Rev A {\bf 68}, 043624 (2003);
D.V. Efremov and L. Viverit, Phys. Rev. B {\bf 65}, 134519 (2002);
L. Viverit, Phys. Rev. A {\bf 66}, 023605 (2002).

\bibitem{composite}
M.Y. Kagan {\it et. al.}, cond-mat/0209481;
M. Lewenstein {\it et. al.}, cond-mat/0306180;
H. Fehrmann {\it et. al.}, cond-mat/0307635.

\bibitem{burnett}
R. Roth and K. Burnett, cond-mat/0310114

\bibitem{xCDW}
H.P. B\"{u}chler and G. Blatter, cond-mat/0304534;
T. Miyakawa \emph{et al.}, cond-mat/0401107

\bibitem{kitaev}
A. Y. Kitaev, cond-mat/0010440

\bibitem{haldane_bf}
F. D. M. Haldane, Phys. Rev. Lett. {\bf 47}, 1840 (1981).

\bibitem{cazalilla}
M.A. Cazalilla, cond-mat/0307033.

\bibitem{general_1D}
A. G\"{o}rlitz, {\it et al.} Phys. Rev. Lett. 
{\bf 87}, 130402 (2001).

\bibitem{solyom} J. Solyom, Adv.  Phys. {\bf 28}, 
201 (1979);
J. Voit, Rep. Prog. Phys., \textbf{58}, 977 (1995).

\bibitem{endnote1} 
In this paper, we use $V_{f/b,\|(\perp)}$ to denote the optical lattice
potential experienced by fermionic/bosonic atoms in the longitudinal
(perpendicular) directions.
Independent tuning of the optical lattices for two
species of atoms can be achieved even with a single pair
of lasers providing the standing beam.
For example, the lattice strengths for bosons and fermions 
in longitudinal direction is given by $V_{b, \|} \sim
\Omega_b^2/\Delta$ and $V_{f, \|} \sim \Omega_f^2/(\delta - \Delta)$,
where $\Delta$ is the detuning of the bosonic state, $\delta$ is the
energy difference between the bosonic and the fermionic state, and
$\Omega_{b/f}$ are the Rabi frequencies, which are propotional to the
laser intensity. By controlling $\Delta$ and the laser intensity,
$V_{b,\|}$ and $V_{f, \|}$ can be varied independently over a wide
range.

\bibitem{ho}
M. A. Cazalilla and A. F. Ho, Phys. Rev. Lett. {\bf 91}, 150403 (2003).

\bibitem{instability}
A.P. Albus, {\it et al.}, cond-mat/0211060;
M.J. Bijlsma, {\it et al.}, Phys. Rev. A {\bf 61}, 053601 (2000). 

\bibitem{Jaksch98}
D. Jaksch {\it et. al.}, Phys.Rev.Lett. {\bf 81}, 3108 (1998).

\bibitem{stringari}
M. Kramer, {\it et. al.},
Phys. Rev. Lett. 88, 180404 (2002).  

\bibitem{halffilling}
In the case of half-filling for
fermions we find that Umklapp processes are
irrelevant (in the renormalization group (RG) sense) outside of the phase
separation region in terms of experimental parameters. Hence, we do not expect a true long range charge
order in the system.

\bibitem{voit_phonon}
J. Voit and Schulz, Phys. Reb. B {\bf 36}, 968 (1987); 
{\it ibid} {\bf 37}, 10068 (1988).

\bibitem{Engelsberg} S. Engelsberg and B.B. Varga , Phys. Rev.  {\bf 136}, 
A1582 (1964).

\bibitem{Definitions} 
We defined $K_{\beta}\equiv e^{2 \theta} \tilde{v}_f (\cos^2\psi/v_A + \sin^2\psi/v_a)$, $K_{\delta}\equiv  K_b v_b (\sin^2\psi/v_A + \cos^2\psi/v_a)$,
$K_{\gamma}^{-1}\equiv  e^{- 2 \theta}/\tilde{v}_f (v_A \cos^2\psi + v_a \sin^2\psi)$,
$K_{\epsilon}^{-1}\equiv  K_b^{-1}/v_b (v_A \sin^2\psi + v_a \sin^2\psi)$,
${K}_{\beta\delta}= e^{\theta} \sqrt{K_b \tilde{v}_f v_b} \sin(2 \psi)/2 (1/v_a - 1/v_A)$, and
${K}_{\gamma\epsilon}^{-1}= e^{-\theta}/\sqrt{K_b \tilde{v}_f v_b} \sin(2 \psi)/2 (v_a - v_A)$. $\psi$ is given by
$\tan 2\psi = 4 \tilde{g} (v_b \tilde{v}_f)^{1/2}/(v_b^2 - \tilde{v}_f^2)$.
Details of the calculations will appear elsewhere.


\bibitem{mott}
A.S. Alexandrov and S.N. Mott, {\it Polarons and Bipolarons}
(World Scientific, Singapore, 1995).

\bibitem{stirring}
C. Raman, {\it et. al.}, Phys. Rev. Lett. {\bf 83}, 2502 (1999);
R. Onofrio, {\it et. al.}, Phys. Rev. Lett. {\bf 85}, 2228 (1999).

\bibitem{kane}
C. Kane and M.P.A. Fisher, 
Phys. Rev. Lett. {\bf 68}, 1220 (1992).

\end{thebibliography}
\end{document}